\documentclass{iaus}
\usepackage{graphicx}

\title[Symbiotic stars as possible progenitors of SNe Ia] 
{Symbiotic stars as possible progenitors of SNe Ia: binary parameters and overall outlook  }

\author[J. Miko{\l}ajewska]   
{J. Miko{\l}ajewska}

\affiliation{Copernicus Astronomical Center, Warsaw, Poland}

\pubyear{2011}
\volume{281}  
\pagerange{1--4}
\date{?? and in revised form ??}
\jname{Proceedings Binary Paths to Type Ia Supernovae Explosions}
\editors{R. Di Stefano  \& M.  Orio , eds.}
\begin{document}

\maketitle

\begin{abstract}
Symbiotic stars are interacting binaries in which the first-formed white
 dwarf accretes and burns material from a red giant companion.
This paper aims at presenting physical characteristics of these objects and discussing their possible link with progenitors of type Ia supernovae. 
\end{abstract}

\section{Introduction}

Symbiotic stars are long-period binaries in which an evolved giant transfers material to a hot and luminous companion surrounded by an ionized nebula. The hot component of the vast majority of symbiotic systems is in fact a hot white dwarf (although in a few cases a neutron star has been found) orbiting close enough to the red giant that it can accrete material from its wind. There are two distinct classes of symbiotic stars: the S-type (stellar) with normal red giants and orbital periods of about 1--15 years, and the D-type (dusty) with Mira primaries usually surrounded by a warm dust shell, and orbital periods generally longer than 10 years. Symbiotic stars are thus interacting binaries with the longest orbital periods, and their study is essential to understand the evolution and interaction of detached and semi-detached binary stars. 
The presence of both accreting white dwarf and the red giant (with its degenerating core) makes also symbiotic binaries a promising nursery for type Ia supernovae (SN Ia), whatever the path to the thermonuclear explosion of CO white dwarf upon approaching or crossing the Chandrasekhar limit (the single degenerate scenario, SD) or by merging of a double white dwarf system (the double degenerate scenario, SD) may be.
The aims if this contribution is to present the basic parameters of these binaries, and discuss their possible link with the progenitors of SN Ia.

\section{Basic characteristics and the nature of symbiotic stars}

The hot components of symbiotic stars have been recently reviewed by Miko{\l}ajewska (2007, 2011). Based on their activity, all symbiotic stars can be divided into two subclasses: ordinary or classical symbiotic star (including Z And, CI Cyg, AX Per, and RW Hya as type examples) and symbiotic novae which are basically thermonuclear novae.

The symbiotic novae (SyNe) are either extremely slow novae, with outbursts going on for decades, or very fast recurrent novae (SyRNe) with very short, $\sim$ several days, timescales and recurrence time of $\sim$ several years or decades.
Such novae are extremely rare in symbiotic systems: among over 200 such systems known, there are only nine slow SyNe, and only four SyRNe. 
There is a strong correlation between the lifetime and luminosity of the novae during their plateau phase as predicted by theoretical studies (see Fig. 6 in Miko{\l}ajewska 2011): the more massive the white dwarf is the higher its luminosity and shorter lifetime on the plateau will be.

The typical (quiescent) hot components of ordinary symbiotic stars are hot, $\sim 10^5$ K, and luminous, $\sim 100$--$10\,000\, \rm L_{\odot}$, and they overlap in the H-R diagram with the region occupied by central stars of planetary nebulae. In most cases, the luminosity is so high that the symbiotic white dwarfs must burn hydrogen-rich material as they accrete. Such a picture is supported by the location of the symbiotic white dwarfs in the luminosity--mass diagram (see Fig. 2 in Miko{\l}ajewska 2011) which generally overlaps with the location of the steady models of Nomoto et al. (2007). The symbiotic white dwarfs, however, appear to be still systematically cooler than the steady models predict (cf. Fig. 1 of Miko{\l}ajewska 2011). 
Among possible explanations of this cool appearance of symbiotic white dwarf is an obscuration by the red giant wind and/or large accretion disk (see also Nielsen 2011, this volume). Most of these hot and luminous symbiotic systems show relatively strong Raman scattered O{\sc vi} which requires significant fraction of the wind to be neutral (e.g. Schmid 1995).

Many of these classical systems, including the prototype Z And, CI Cyg, and AX Per, show occasional 1--3 mag optical eruptions with timescales from months to years, during which the hot component maintains roughly constant luminosity whereas its effective temperature decreases to $\sim 10^4$ K. The mechanism of this activity has been only recently explained by unstable disk-accretion onto hydrogen-shell burning white dwarf (Miko{\l}ajewska 2003, Sokoloski et al. 2006), and it is still waiting for a quantitative model.

It is interesting that the hot components of the best studied SyRNe, RS Oph and T CrB, show intrinsic activity (high and low states) between their TNR nova outbursts, with timescales, and other features (e.g. a blue continuum with A/F-type absorption lines) similar to those characterizing the multiple outburst activity of Z And-type stars, and very likely due to some accretion disk instability (Gromadzki et al. 2008). 

Finally, some symbiotic white dwarfs with luminosities of $\sim 10$--$100\, \rm L_{\odot}$ which reveal optical flickering and hard X-ray emission indicating the white dwarf being massive, seem to be accretion-powered (e.g. Luna \& Sokoloski, 2007; Mukai 2011, this volume; Sokoloski 2011, this volume). At least some of these objects, e.g. CH Cyg, EG And, RT Cru, also show intrinsic activity similar to the Z And-type systems and SyRNe between nova outbursts. 

The orbital periods are presently known for 92 systems, all but R Aqr belonging to the S-type (Miko{\l}ajewska 2011; Gromadzki et al. 2011, in preparation). About 70\,\% of the systems have $P_{\rm orb} \sim 400$--1000 days and only $\sim$ 20\,\%, above 1000 days.
The orbital periods for the four SyRNe are all below $\sim 600$ days whereas the slow SyNe have $P_{\rm orb} \gtrsim 800$ days, and they fall on the shorter and longer period tail of the S-type systems, respectively.
The spectroscopic orbits for the cool giant  are known for 37 systems, and for about 20 of them, including two SyRNe, 
the mass ratios have been also estimated (Miko{\l}ajewska 2011, and references therein).
The cool giant masses peak around $1.6 \, \rm M_{\odot}$, and most white dwarfs have masses between 0.4--$0.8\, \rm M_{\odot}$. There is, however, interesting difference between the SyRNe and the remaining systems: in both RS Oph and T CrB, the giant is the less massive component, with mass below $1\, \rm M_{\odot}$, whereas their massive white dwarfs, with mass of $\sim 1.1$--$1.4\, \rm M_{\odot}$, are promising candidates for the SN Ia progenitors.

Although it is usually assumed that symbiotic stars interact via stellar wind rather than Roche lobe overflow (RLOF), there is an increasing number of systems with ellipsoidal changes in their curves. So far, such changes have been detected in  $\sim 30\,\%$ of the systems with orbital periods below $\sim 1000$ days, including the three SyRNe (Miko{\l}ajewska 2007; Schaefer 2009; Gromadzki et al. 2011, in preparation). Such changes can be, however, detected only in systems with relatively high inclination, and, in addition, they usually show up only in the red and near-IR light where the cool giant dominates the spectrum, whereas for most systems only the optical light curves are available. So, RLOF and high mass transfer rates, $\sim \rm a\,few \times 10^{-8}$--$10^{-7}\, \rm M_{\odot}/yr$ can be very common in symbiotic stars with shorter orbital periods. 

The presence of tidally distorted donors and RLOF is especially important in relation to symbiotic stars as possible progenitors of SN Ia, because the resulting high mass transfer rate allows the white dwarf to gain significant portion of the red giant envelope. This is particularly promising in the case of SyRNe where the white dwarf is already very massive, and although the mass of the red giant envelope is below $0.5\, \rm M_{\odot}$, such an efficient accretion should let the white dwarf to grow to the Chandrasekhar limit.

The SyRNe are, however, not the only promising progenitors of SN Ia among the symbiotic stars.

\section{V407 Cyg: a connection to progenitor of Kepler's supernova?}

The most recent symbiotic nova, V407 Cyg, shares many similarities with RS Oph and other SyRNe (e.g. Miko{\l}ajewska 2011, and references therein). In particular, its very fast outburst development indicates that the white dwarf is as massive as in the SyRNe. Moreover, prior to its 2010 nova eruption,  V407 has been known  as wind-accreting D-type symbiotic system, with high and low states similar to those observed in the SyRNe, and other (like CH Cyg and R Aqr) accretion-powered systems.  

Recently, Chiotellis (2011, this volume) has argued that the main features of the present remnants of Kepler's supernova are consistent with a symbiotic binary
in which the white dwarf was accreting mass from the strong wind of a 4--5\,$ \rm M_{\odot}$ AGB star.  
Although a typical D-type symbiotic system seems to contain a low-mass Mira-type variable (e.g. Gromadzki \& Miko{\l}ajewska 2009), V 407 Cyg is exceptional in the sense that it has the longest pulsation period among the symbiotic Miras and the only one which is Li-rich (Tatarnikova et al. 2003). Such a Li enrichment is very rare among evolved giants, and it is usually accounted for as being due to hot bottom burning occurring in stars with initial masses of 4--8\, $\rm M_{\odot}$. 
Such a Mira can also have a strong wind, $\sim 10^{-5}\, \rm M_{\odot}/yr$, enough to power its massive white dwarf companion with accretion rate at least occasionally reaching the high rates of the SyRNe.

\section{AR Pav: a massive white dwarf accreting from tidally distorted giant}  

AR Pav is a symbiotic binary with the orbital period of 605 days consisting of a $\sim 2.5\, \rm M_{\odot}$ M6 giant transferring mass to an accretion disk surrounding a $\sim 1\, \rm M_{\odot}$ companion (Quiroga et al. 2002). The companion is a hot, $\sim 10^5\, \rm K$ and luminous, $\sim 5000$--$10\,000\, \rm L_{\odot}$, white dwarf more or less steadily burning H-rich material accreted from the tidally distorted giant. The near-IR light curves require the giant to fill or nearly fill ($\sim 96\, \%$) its Roche lobe (Rutkowski et al. 2007) whereas an accretion rate of $\sim  10^{-7}\, \rm M_{\odot}/yr$ is required to power the observed hot component luminosity via nuclear burning.
The hot component has remained relatively active over last decades, and the optical light curves show, in addition to eclipses, a moving bump similar to that observed in other active Z And-type systems and RS Oph between its nova outbursts (Gromadzki et al. 2008), pointing to some accretion disk instability in this case as well. 

The further evolution of AR Pav is very interesting: whether the RLOF will remain stable and eventually become a system like the SyRNe, RS Oph and T CrB, or it is just about to enter a common envelope (CE) phase, after which a close double $\sim 1\, \rm M_{\odot} + 0.5\, \rm M_{\odot}$ white dwarf (DD) system will be formed.

AR Pav is not the only classical symbiotic star with RLOF; at least 30\% 
of those with orbital periods below $\sim 1000$ days contain tidally distorted giants, including both stable and Z And-type multiple outburst systems. 
They all contain white dwarfs with masses between 0.5--1\,$\rm M_{\odot}$ and 1--3\,$\rm M_{\odot}$ red giant companions with core masses of $\sim 0.45$--$0.55\, \rm M_{\odot}$. 
At least some of them may either end as RS Oph-type systems or after CE phase will become close pairs of white dwarfs with total mass comparable to the Chandrasekhar limit.

\section{Summary}

Understanding binary interaction and evolution of symbiotic stars is essential to solving the SN Ia progenitor problem.

In particular, symbiotic stars contain white dwarfs efficiently accreting and, in most cases, steadily burning H-rich matterial allowing them to grow in mass. 
Some are able to produce high mass white dwarfs.  The white dwarf mass is already close to the Chandrasekhar limit in SyRNe. A very massive white dwarf accreting at a high, $\sim 10^{-7}\, \rm M_{\odot}/yr$, rate from its Mira-type companion is present in V 407 Cyg. Very promising are also systems like AR Pav with Roche lobe filling giants transferring material to, relatively massive in some cases, white dwarf companions, which, if they will pass through a CE phase, may become close pairs of DD with total mass of about the Chandrasekhar mass.
Summarizing, some symbiotic systems appear to be promising candidates for the SN Ia for both SD and DD scenarios. 

\begin{acknowledgments}
I gratefully acknowledge the financial support from the LOC through the IAU travel grant.
\end{acknowledgments}

\end{document}